\documentclass[12pt,a4paper]{article}
\usepackage[utf8]{inputenc}
\usepackage{amsmath}
\usepackage{graphicx}
\usepackage{amssymb}
\usepackage{amsthm}
\usepackage{color}

\usepackage{lineno}

\usepackage{mathtools}

\usepackage[breaklinks=true,backref=page]{hyperref}

\usepackage[english]{babel}
\usepackage{natbib}

\usepackage[nottoc]{tocbibind}

\usepackage{multirow}

\usepackage{authblk}

\usepackage{amsthm}
\theoremstyle{plain}
\newtheorem{thm}{Theorem}[subsection]

\newtheorem{prop}[thm]{Proposition}

\theoremstyle{definition}

\theoremstyle{remark}

\newcommand{\minitab}[2][l]{\begin{tabular}{#1}#2\end{tabular}}

\title{Mathematical model of mating probability and fertilized egg production in helminth parasites}

\author[1,2]{\small Gonzalo Maximiliano LOPEZ \thanks{gonzalo.maximiliano.lopez@gmail.com}}
\author[1]{\small Juan Pablo APARICIO \thanks{juan.p.aparicio@gmail.com}}
\affil[1]{\small Instituto de Investigaciones en Energ\'ia no Convencional, Consejo Nacional de Investigaciones Cient\'ificas y T\'ecnicas,
Universidad Nacional de Salta, Av. Bolivia 5150, 4400 Salta, Argentina }
\affil[2]{\small Departamento de Matem\'atica, Facultad de Ciencias Exactas, Universidad Nacional de Salta, Av. Bolivia 5150, 4400 Salta, Argentina}

\date{}

\begin{document}

\maketitle

\begin{abstract}

In the modeling of parasite transmission dynamics, understanding the reproductive characteristics of these parasites is crucial.	

This paper presents a mathematical model that explores the reproductive behavior of dioecious parasites and its impact on transmission dynamics.

Specifically, the study focuses on the investigation of various reproductive variables such as the mating probability and the fertilized egg production in the case of helminth parasites.

While previous studies have commonly assumed Poisson and negative binomial distributions to describe the distribution of parasites among hosts, this study adopts an arbitrary distribution model and examines its consequences on some reproductive variables. These variables include mean number of fertile females, mean egg production, mating probability and mean fertilized egg production. 

In addition, the study of these variables takes into account the sex distribution of the parasites and whether male and female parasites are considered to be distributed together or separately.



We show that the models obtained for the case of male and female parasites distributed separately in the hosts are ecologically unrealistic.

We present the results obtained for some specific models and we tested the models obtained in this work using Monte Carlo simulations.

\end{abstract}

\tableofcontents
	\section{Introduction}
	
	The mating probability of female parasites is a key factor in understanding the transmission dynamics of dioecious parasites. 
	It is influenced by several factors, including parasite mating habits (polygamous vs. monogamous), sex ratio, mean parasite burden (mean of number of parasites per host), and the distribution pattern of female and male parasites within the host population (parasite sex distributed together or separately).
	
	Previous models \citep{macdonald1965dynamics,naasell1973transmission} estimated the mating probability by assuming that adult parasites are independently and randomly distributed among the hosts (Poisson distribution).

	Later models \citep{may1977togetherness,bradley1978consequences,may1993biased} estimated mating probability by assuming that adult parasites are aggregated or clumped in their distribution (negative binomial distribution).
	
	Theoretical models suggest that a high mating probability is associated with polygamous mating, a male-biased sex ratio, a high mean parasite burden, and a high degree of aggregation, provided that male and female parasites aggregate together \citep{may1977togetherness,bradley1978consequences,may1993biased}.
	Mating probabilities for different types of polygamous parasites (ectoparasites, nematodes and helminths) have been calculated using previous models \citep{haukisalmi1996variability,cox2017mate}.
	
	Another important aspect of parasite transmission dynamics is the production of new parasite stages, such as infective eggs and larvae, which is influenced by female parasite mating and female fecundity \citep{anderson1992infectious}. 
	
	In the case of parasitic helminths in particular, female fecundity exhibits negative density-dependent processes that limit growth rates in high-density populations and help to stabilise natural communities. 
	
	This has important implications for the stability and transmission dynamics of these populations, and it is essential to incorporate these processes into mathematical models with high accuracy to better understand their dynamic behaviour \citep{anderson1978regulation,anderson1992infectious,churcher2005density,churcher2006density,churcher2008density}. 
	In particular, density-dependent egg production has been reported in human helminth infections, with per capita egg production decreasing as the number of parasites per host (parasite burden) increases \citep{hall2000geographical,churcher2006density}. 
	
	In this paper, we extend previous mating probability models to a more general and challenging scenario where the parasite distribution in the host population follows an arbitrary statistical model. 
	This extension is necessary because several studies \citep{abdybekova2012frequency,crofton1971quantitative,denwood2008distribution,lopez2023simple,ziadinov2010frequency}
	have reported parasite distributions other than the traditional ones, such as zero-inflated and hurdle models.
	
	
	Based on this new model for the mating probability of female parasites, and taking into account the density-dependent aspects of helminth parasite fecundity, our second aim in this work is to determine the mean number of infective egg production per host and 
	the probability of producing infective eggs. 
	We also show how these variables can be incorporated into mean burden-based models of helminth infection. 
	
	Finally, we tested the models obtained in this work using Monte Carlo simulations in which we recreated some distributions of parasites in the host population.

	\section{Parasite sex distribution among hosts}
	\label{sec:probapareamiento}
	
%
%
%
%
%

\subsection{Distribution and abundance of parasites}

The distribution of parasites among hosts is a crucial aspect of host-parasite interactions. 
Typically, parasites exhibit a clustered or aggregated distribution, where a small proportion of hosts harbor the majority of parasites, while the majority of hosts remain parasite-free \citep{shaw1995patterns,shaw1998patterns}. 
This distribution follows the widely recognized 20-80 rule, where 20\% of individuals contribute to 80\% of the parasite burden \citep{woolhouse1997heterogeneities}. 

Aggregation is considered a fundamental characteristic of parasitism and is often referred to as the "First Law of Parasitism" \citep{crofton1971quantitative,poulin2007there}. 

In statistical terms, a distribution is considered aggregated when the variance-to-mean ratio of the number of parasites per host is significantly greater than one. To describe the distribution of parasites among hosts, the negative binomial distribution is commonly employed. 
This discrete and flexible distribution, characterized by two parameters \citep{fisher1941negative}, fits well with the observed aggregation in nature \citep{bliss1953fitting}. 
Consequently, several theoretical models of host-parasite dynamics implicitly assume the negative binomial distribution \citep{anderson1978regulation,adler1992aggregation}. 

However, it is important to note that the negative binomial distribution provides only a phenomenological description of aggregation, and its parameters are not explicitly linked to underlying processes governing host exposure to parasites and parasite success in infecting hosts \citep{duerr2000stochastic,gourbiere2015fundamental}.

In parasitology, the sex ratio of a parasite population is often quantified by the proportion of female parasites ($\alpha$) and the proportion of male parasites ($\beta=1-\alpha$). The male-to-female sex ratio, expressed as $\alpha / \beta : 1$, serves as a significant metric in the study of parasitic infections. 
Another important metric is the mean number of female parasites per host. If $m$ represents the mean parasite burden, the mean number of female and male parasites can be calculated as $\alpha m$ and $\beta m $, respectively. By understanding the sex ratio and the mean number of female parasites per host, researchers can gain valuable insights into the transmission and spread of parasitic infections.

\subsection{Female and male parasites distributed together} 
\label{sec:distsexo}

In this section we use random variables to model the distribution of parasites in hosts. 
Let $W$ be the random variable representing the number of parasites in a host, also known as the parasite burden. 
The variable $W$ is distributed according to the distribution of parasites in the hosts.
In this section we assume that male and female parasites are distributed together in hosts.
Therefore, we consider the number of female and male parasites per host as two dependent random variables, denoted $F$ and $M$, respectively.
The total number of parasites per host, $W$, is then the sum of $F$ and $M$. 
To model the distribution of female parasites per host, we can model the number of female parasites per host, $F$, as a random sum of other random variables, modeled by a stopped-sum model
\citep{johnson2005univariate}, defined as $F=\sum_{i=1} ^W X_i$ where the $X_i$ are independent identically distributed (iid) Bernoulli random variables with parameter $\alpha$ ($X_i\sim \mathrm{Ber}(\alpha)$).
Its probability generating function (pgf) is the function composition   
$G \circ G_B$, where $G$ is the pgf of the random variable $W$ and $G_B$  is the pgf of the Bernoulli distribution given by $G_B(s)=\beta + \alpha s$.
Therefore, its pgf $G_F$ is of the form 
	\begin{equation}\label{genf}
	\begin{split}
	G_F(s)=&G \circ G_B(s)=G(\beta + \alpha s)\\
	=&\sum_{n\geq 0}\sum_{j=0}^{n} \mathrm{Pr}(W=n)\binom{n}{j}\alpha^j\beta^{n-j}s^j
	\end{split}
	\end{equation}
	The first moments of $F$ are
	\begin{equation}
	\mu_F=\alpha \mu \qquad \sigma_F^2=\alpha^2\sigma^2+ \alpha\beta\mu,
	\end{equation}
	where $\mu$ and $\sigma^2$ are the mean and variance of $W$. 
	The index of dispersion or variance-to-mean ratio \citep{cox1966statistical}
	, $D=\frac{\sigma_F^2}{\mu_F}$, is given by
	\begin{equation}
		\label{vmr}
		D=\alpha\frac{\sigma^2}{\mu}+\beta
	\end{equation} 
	where $\frac{\sigma^2}{\mu}$ 
	is the variance-to-mean ratio of $W$.
	Therefore, if $W$ is over-dispersed, so will $F$.

	Similarly, we define the random variable $ M $, the number of male parasites per host, as a randomly stopped-sum definity by $ M = \sum_{i = 1}^W Y_i $ where the $ Y_i = 1-X_i $ are iid Bernoulli random variables with $ Y_i\sim \mathrm{Ber}(1-\alpha)$.	
	Its pgf is of the form $G_M(s)=G(\alpha + \beta s)$
	and the mean and variance are $\mu_M=\beta\mu$ and $\sigma_M^2=\beta^2\sigma^2+ \alpha\beta\mu$, respectively.	
	Note that for definition  $F$ and $M$ are dependent random variables.

	In practice, the distribution of the variable $W$ is known.
	Therefore, by obtaining the pgf $G_F$ and $G_M$, we can determine the distributions of the variables $F$ and $M$. 
	A general class of distributions exists in which the statistical model of $W$ is conserved. 
	This class includes the Poisson, binomial and negative binomial distributions, among others, which are widely used for counting models \citep{johnson2005univariate}.

	In the paper, we define $G(s;\theta)$ by $\sum_{n\geq 0}p(n;\theta)s^{n}$, where $p$ is the probability mass function of the parasite distribution and $\theta$ is a parameter vector in which $\theta$ can include the mean, $m$, of the parasite distribution.
	
	

	\subsection{Fertilized female parasites and mating probability}
	
	In this paper we considered a polygamous mating system for parasites, i.e. a male parasite can fertilize all female parasites in the same host \citep{may1993biased}. 
	Since the expression $\sum_{j=0}^{n-1} p_n\binom{n}{j}\alpha^j\beta^{n-j}$
	is the probability of having at least one male parasite in a parasite burden of size $n$, we obtained the following result
	\begin{prop}\label{hembrasfecun}
		The mean number of fertilized female parasites is given by     
		\begin{equation}
		\alpha  m - \alpha G'(\alpha).
		\end{equation}
	\end{prop}

	Then we can estimate the mating probability of a female parasite 
	as the quotient of the mean number of fertilized female parasites per host, $\alpha m -\alpha G'(\alpha)$, 
	and the mean number of female parasites per host, $\alpha m$. 
	Therefore, a expression for mating probability (denoted by $\Phi$) as a function of mean parasite burden $m$ is given by
\begin{thm} The mating probabillity of a female parasite is given by
	\begin{equation}\label{probrepro1}
		\Phi(m)
		=1-\frac{G'(\alpha;m)}{m}.
	\end{equation}
\end{thm}
	\subsection{Density-dependent fecundity of helminth parasites}
	In population ecology, density-dependent processes occur when population density affects growth rates. In the case of parasites, these processes can affect their fecundity, establishment and survival in the host.
	For example, in helminth parasites, their fecundity has been observed to decrease as the parasite burden in the host increases \citep{churcher2006density,walker2009density}. 
	This phenomenon is known as density-dependent fecundity, and it is explained by a negative exponential function that relates per capita fecundity to parasite burden
	\begin{equation}
		\lambda(n)=\lambda_0 \exp[-\gamma(n-1)],
	\end{equation} 
	where $\lambda(n)$ is the per capita female fecundity within a host with a parasite burden of size $n$, $\lambda_0$ is the intrinsic fecundity in the absence of density-dependent effects, and $\gamma$ is the density-dependent intensity.
	To simplify the notation in the rest of the text, we will express female fecundity by $\lambda(n)=\lambda_0 z^{n-1}$ where $z=e^{-\gamma}$.
	In summary, density-dependent effects are an important factor in parasite ecology, and may have significant implications for parasite population dynamics. For more information on this subject, see the study by \citet{hall2000geographical} on \textit{Ascaris lumbricoides}.


	\subsubsection{Fertilized egg production and egg fertility probability (mating probabillity with density-dependent effects)}
	Due to the effects of density-dependent fecundity, egg production per female decreases as the parasite burden in the host increases. 
	Therefore, if $j\lambda(n)$ is the egg production of $j$ females within a host with $n$ parasites, and $p_n\binom{n}{j}\alpha^j\beta^{n-j}$ is the probability that a host with $n$ parasites has $j$ females. 
	Then, we obtain the following result for the mean egg production per host
%
	\begin{prop}\label{prodhuevos}
		The mean egg production per host is given by
		\begin{equation}
		\lambda_0\alpha G'(z).
		\end{equation}
	\end{prop}


	For the case of the mean fertilized egg production per host, we use the previous proof, but considering only the egg production of fertilized female parasites. 
	As a result, an expression for the mean fertilized egg production per host is given by
	\begin{prop}\label{prodhuevosfecun}
		The mean fertilized egg production per host is given by
		\begin{equation*}
		\lambda_0\alpha G'(z)\left[1-\frac{G'(\alpha z)}{G'(z)} \right].
		\end{equation*}
	\end{prop}

According to the results obtained previously, if we consider the quotient of the mean fertilized egg production and the mean egg production, we obtain the egg fertility probability or the mating probability of the female parasites under the density-dependent effects.
Therefore, the mating probability with density-dependent effects of a female parasite as a function of the mean parasite burden $m$ is given by


	\begin{thm} The mating probabillity with density-dependent effects of a female parasite is given by
	\begin{equation}\label{probrepro2}
	\phi(m)=1-\frac{G'(\alpha z;m)}{G'(z;m)}. 
	\end{equation}
	\end{thm}
	From this expression \eqref{probrepro2} we see that for the case where there is no density-dependence ($z \approx 1$) this expression is equivalent to the expression \eqref{probrepro1}, so this is a generalization of the mating probability, $\Phi$, obtained above.
	
	\subsubsection{An application for mean burden-based models for helminth infections}
	In deterministic population models based on the mean parasite burden for the transmission dynamics of helminth infections, such as \citep{anderson1985helminth,anderson1992infectious,lopez2022modeling,truscott2014modeling}, it is necessary to know the effective transmission contribution of the female population to the parasite reservoir (in form of eggs or larvae), assuming density-dependent processes (positive and/or negative) within the parasite life cycle. The effective transmission contribution term is commonly denoted by $\psi$ and can be calculated as shown in \citep{churcher2005density,churcher2006density,lopez2022modeling},
	\begin{equation}\label{psi}
	\psi=\frac{\sum_{n\geq 0}\sum_{j=1}^{n}j\lambda(n)p_n\binom{n}{j}\alpha^j\beta^{n-j}}
	{\sum_{n\geq 0}\sum_{j=0}^{n}jp_n\binom{n}{j}\alpha^j\beta^{n-j}}
	\end{equation}
	where the density-dependent fecundity $\lambda(n)$ is redefined as $\lambda(n)/\lambda_0$.
	Therefore, the function $\psi$ has a maximum value of one and separates the density-independent term $\lambda_0$ from the density-dependent processes.
	
	Using the results obtained in this paper, we can calculate $\psi$ as a function of the mean parasite burden $m$, as follows
	\begin{equation}\label{psi}
	\psi(m)
	=\frac{G'(z;m)}{m}   
	\end{equation}
	Therefore, if we know the distribution of parasites in hosts, 
	we can calculate the mean egg production per host as
	\begin{equation}
	\lambda_0\alpha m \psi(m)=\lambda_0 \alpha G'(z;m) 
	\end{equation} 
	However, only hosts with at least one female and one male parasite will effectively contribute to the parasite reservoir by producing fertilized (or infective) eggs. The mean fertilized egg production per host is then (see e.g. \citet{anderson1992infectious,lopez2022modeling})
	\begin{equation}
	\lambda_0\alpha m \psi(m) \phi(m)= \lambda_0 \alpha G'(z;m) \left[1-\frac{ G'(\alpha z;m)}{G'(z;m)}\right] 	
	\end{equation}
	where we assume that $\psi$ and $\phi$ are functions of the mean parasite burden $m$.

	\subsection{Some examples}
	\label{ss:examples-dep}
	In this section we will consider the most common statistical models used to describe the distribution of parasites among hosts.
	\subsubsection{Poisson}
	\label{ss:po-dep}
	A simple model for the distribution of parasites per host  is the Poisson distribution \citep{lahmar2001frequency,macdonald1965dynamics},
	\begin{equation}
	\mathrm{Pr}(X=x)=\frac {m^{x}e^{-m}}{x!},
	\end{equation}     
	where $m$ is the mean parasite burden and its pgf is given by
	\begin{equation}
	\begin{split}
	G(s)&=e^{m(s-1)}.
	\end{split}
	\end{equation}
	For this parasite distribution 
	the effective transmission contribution of female parasites to the transmission cycle is given by (see eq \eqref{psi})
	%
	\begin{equation}
	\psi(m)=e^{m(z-1)}.
	\end{equation}
	Other important factors in parasite dynamics are the mating probability $\Phi$ and the mating probability with density-dependent effect $\phi$, which are given by (see eq \ref{probrepro1},\ref{probrepro2}).
	%
	\begin{equation}
	\Phi(m)=1-e^{-m\beta}, \qquad
	\phi(m)=
	1-e^{-mz \beta}.
	\end{equation}
	The expression for the mating probability $\Phi$ is the same as in \citep{anderson1992infectious,may1993biased,may1977togetherness}.
	The expression $\phi$ generalizes the mating probability of these works for the case of helminth parasites.

	

	\subsubsection{Negative binomial}
	\label{ss:nb-dep}
	In most cases, soil-transmitted helminths, present a distribution of parasites per host that can be well described by a negative binomial distribution \citep{bundy1987epidemiology,hoagland1978necator,seo1979frequency},
	\begin{equation}
	\mathrm{Pr}(X=x)=\frac{\Gamma(k+x)}{\Gamma(x+1)\Gamma(k)}\left( \frac{k}{k+m}\right) ^k \left( \frac{m}{k+m}\right) ^x,
	\end{equation}
	where $m$ is the mean parasite burden and $k$ is the inverse dispersion parameter of the parasites. The corresponding pgf is given by
	\begin{equation}
	\begin{split}
	G(s)&=\left[ 1-\frac{m}{k}(s-1)\right] ^{-k}.\\
	\end{split}
	\end{equation}
	Therefore, 
	the expression for $\psi$, the effective transmission contribution, which is given by (see eq. \eqref{psi})
	\begin{equation}\label{phinb-dep}
	\psi(m)=\left[ 1-\frac{m}{k}(z-1)\right] ^{-(k+1)}.
	\end{equation}     
	Finally, the mating probability, $\Phi$, and the mating probability with density-dependent effect, $\phi$, are given by (see eq. \eqref{probrepro2})
	\begin{equation} 
	\Phi(m)=
	1-\left[  1+\frac{m\beta}{k}\right]  ^{-(k+1)}, 
	\qquad
	\phi(m)=
	1-\left[ \frac{ 1-\frac{m}{k}(\alpha z-1)}{1-\frac{m}{k}(z-1) }\right]  ^{-(k+1)}. 
	\end{equation}
	The expression of $\Phi$ is the same as in \citet{anderson1992infectious,may1993biased,may1977togetherness} and the expression $\phi$ results in a generalization of the mating probability for the case of helminth parasites.
	
		
	\subsubsection{Zero-inflated and hurdle models}
	Other commonly used models are the zero-inflated and hurdle models (see e.g. \citet{abdybekova2012frequency,crofton1971quantitative,denwood2008distribution,ziadinov2010frequency,lopez2023simple}).
	For a zero-inflated model, its probability mass function is
	\begin{equation*}\label{zid}
	\mathrm{Pr}(Y=y)= \left\{ \begin{array}{lc}
	\pi + (1-\pi)p_0 & y=0 \\
	\\ (1-\pi)p_y  & y\neq 0
	\end{array}
	\right.
	\end{equation*}
	where $p$ is the probability mass function of a distribution without excess of zero counts and $G_X$ is the corresponding pgf. Then the pgf of the zero-inflated distribution is
	\begin{equation*}
	G_Y(s)=\pi+(1-\pi)G_X(s),
	\end{equation*}
	and the mean burden is  
		\begin{equation*}
	m_Y=(1-\pi)m_X.
	\end{equation*}
	For this model the 
	expression for $\psi$, the mean contribution per female parasite, which is given by
	\begin{equation}\label{zipsi}
	\psi= \frac{G_Y'(z)}{m_Y}=  \frac{(1-\pi)G_X'(z;m_X)}{m_Y}=\frac{G_X'\left( z;\frac{m_Y}{1-\pi}\right) }{\frac{m_Y}{1-\pi}}. 
	\end{equation}
	Finally the mating probability $\phi$ can be calculated by     
	\begin{equation}\label{ziphi}
	\phi=1-\frac{G_Y'(\alpha z)}{G_Y'(z)}=1-\frac{G_X'\left(\alpha z;\frac{m_Y}{1-\pi}\right)}{G_X'\left( z;\frac{m_Y}{1-\pi}\right)}. 
	\end{equation}
	
	A hurdle model is a two-part model, the first part, $\pi$, is the probability of observing the zero value, and the second part is the probability of observing non-zero values. 
	The use of hurdle models is often motivated by an excess of zero counts in the data, which is not sufficiently accounted for in more standard statistical models \citep{johnson2005univariate}. 
	For this model, its probability mass function is given by
	\begin{equation*}\label{hd}
	\mathrm{Pr}(Y=y)= \left\{ \begin{array}{lc}
	\pi & y=0 \\
	\\ (1-\pi)\frac{p(y)}{1-p_0}  & y\neq 0
	\end{array}
	\right.
	\end{equation*}
	Its pgf $G_Y$ and its mean are of the form
	\begin{equation*}
	\begin{split}
	G_Y(s)&=\pi+(1-\pi)\frac{G_X(s)-p_0}{1-p_0},\\
	m_Y&=(1-\pi)\frac{m_X}{1-p_0}.
	\end{split}
	\end{equation*}
	Therefore, the expresions for $\psi$ and $\phi$ are given by
	\begin{equation}
	\label{hm}
	\begin{split}
	\psi&= \frac{G_Y'(z)}{m_Y}= \frac{\rho G_X'(z;m_X)}{ m_Y} =\frac{G_X'\left( z;\frac{m_Y}{\rho}\right) }{\frac{m_Y}{\rho}},\\
	\phi&=1-\frac{G_Y'(\alpha z)}{G_Y'(z)}=1-\frac{G_X'\left( \alpha z;\frac{m_Y}{\rho}\right) }
	{G_X'\left( z;\frac{m_Y}{\rho}\right) } .
	\end{split}
	\end{equation}
	where $\rho=\frac{1-\pi}{1-p_0}$.
	
	In practice, we need to use the expressions on the right-hand side of the equations \eqref{zipsi}\eqref{ziphi}\eqref{hm} to obtain the $\psi$ and $\phi$ expressions for the zero-inflated and hurdle models. 
	Some examples of this are given below.
	\subsubsection{Zero-inflated Poisson and zero-inflated negative binomial models}
	\label{ss:zim-dep}
	Negative binomial distribution is widely used to describe parasite distribution in hosts \citep{crofton1971quantitative,seo1979frequency}.
	However, in many cases the negative binomial distribution (or other similar distributions) cannot account for the observed excess of zero counts \citep{crofton1971quantitative,lopez2023simple}. 
	One solution to this problem is zero-inflated models, which have been widely used for parasite counts in the last decade \citep{abdybekova2012frequency,denwood2008distribution,lopez2023simple,ziadinov2010frequency}.
	
\begin{figure}[h!]
		\centering 	\includegraphics[width=0.6\linewidth]{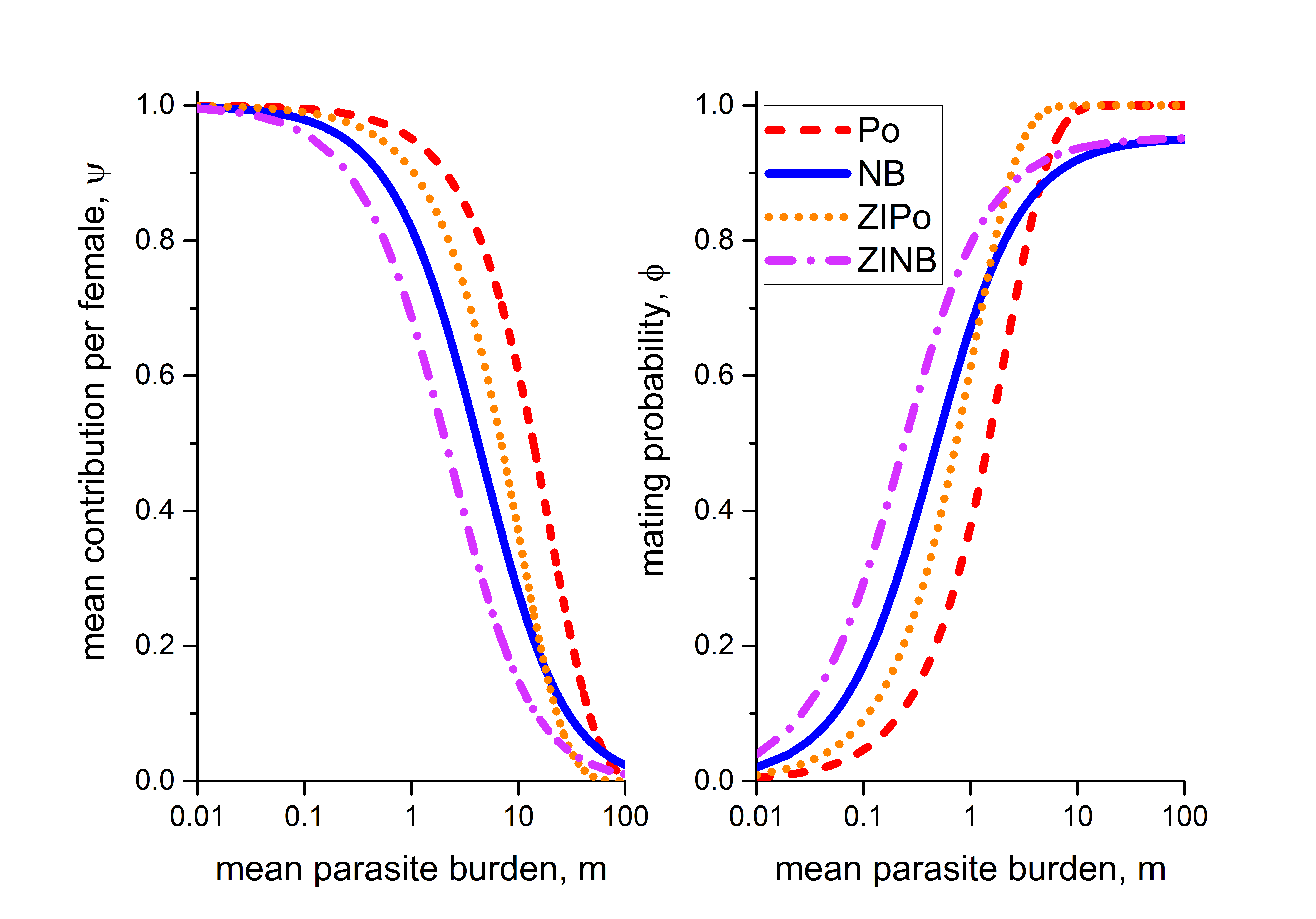}
		\caption{
		The mean effective contribution per female parasite, $\psi$ (left), and the mating probability, $\phi$ (right), corresponding to Poisson (dashed curve), negative binomial (solid curve), zero-inflated Poisson (dotted curve) and zero-inflated negative binomial (dashed dot curve) distributions. All as a function of mean parasite burden $m$.}
		\label{fig:phi}
	\end{figure}

	
	In table \ref{table:function}  we present the expressions for the effective transmission contribution and the mating probability with density-dependent effect for the zero-inflated Poisson and zero-inflated negative binomial models.
	\begin{table}[h!]
		\caption{The effective transmission contribution, $\psi$, and the mating probability, $\phi$, for zero-inflated Poisson (ZIPo) and zero-inflated negative binomial (ZINB) models.}
		\label{table:function}
		\centering
		\vspace{0.3cm}
		\resizebox{\textwidth}{!}{
			\begin{tabular}{c c c}
				\hline  
				\multirow{2}{*}{\minitab[c]{Statistical \\ model}}    & \multirow{2}{*}{ $\psi$} & \multirow{2}{*}{$\phi$}\\ 
				&  &\\
				\hline
				&  &\\
				$\mathrm{ZIPo}(\pi,\frac{m}{1-\pi})$  & $\exp\left( \tfrac{m}{1-\pi}(z-1)\right) $ & $1-\exp\left( -\tfrac{m z \beta}{1-\pi}\right) $	\\ 
				&  &\\ 
				$\mathrm{ZINB}(\pi,\frac{m}{1-\pi},k)$  & $\left[ 1-\tfrac{m}{k(1-\pi)}(z-1)\right] ^{-(k+1)}$ & $1- \left[ \dfrac{1-\tfrac{m}{k(1-\pi)}(\alpha z-1)}{1-\tfrac{m}{k(1-\pi)}(z-1)}\right] ^{-(k+1)}$	\\ 
				&  &\\
			\end{tabular} 
		}
	\end{table}
	Note that to obtain the expression for the mating probability $\Phi$, we have to replace the variable $z$ in the expression for $\phi$ with 1.
	Plots of the effective transmission contribution ($\psi$) and the mating probability ($\phi$) for all the distributions discussed above are shown in Figure \ref{fig:phi}. 
	We consider the parameters $\alpha=0.574$, $k=0.3$, $z=0.92$, $\pi=0.3$ (\cite{anderson2014coverage,seo1979egg}).

\newpage		
\section{Can female and male parasites be distributed separately?}
\label{sec:disindep}


In this section we consider a case where the male and female parasites are distributed separately in the hosts. We assume that the random variables representing the number of male and female parasites per host are independent.

We show that in the common case of over-dispersed distributions, such as the negative binomial, independence leads to counter-intuitive results. This tells us that infection processes capable of producing independent distributions are unrealistic.

In the work of \citep{may1977togetherness,bradley1978consequences}, studies of the probability of mating were carried out taking into account a special condition: the separate distribution of female and male parasites in the hosts. 

They found that the probability of mating is lower when considering this separate distribution than when considering the joint distribution. This is because the distribution of parasites is over-dispersed, making it difficult to find mates when there are few hosts with many parasites of the same sex.

On the other hand, empirical tests of mating probability have been carried out by \citep{cox2017mate,poulin2007there}, assuming special conditions where female and male parasites are distributed separately in hosts. These studies concluded that the probability of mating is higher when the parasites are distributed together rather than separately.


\subsection{Parasite sex distribution}
	
	In the previous section, we introduced the random variable $W$, which represents the number of parasites in a host. We also introduced two additional random variables, $F$ and $M$, representing the number of female and male parasites per host, respectively.
	
	In this section we focus on the case where male and female parasites are distributed separately among hosts. Therefore, we assume that the random variables $F$ and $M$ are independent. This assumption allows us to investigate and verify the following properties:

	
	
	
	\begin{equation}\label{independencia}
	\begin{split}
	W&=F+M\\
	G(s)&=G_F(s)G_M(s)
	\end{split}
	\end{equation}
	where $G$, $G_F$ and $G_M$ are probability generating function of the variables $W$, $F$ and $M$, respectively.
	From the definition of these random variables, we obtain that the first moments of the variables $F$ and $M$ are respectively
	\begin{equation}\label{indep-moments}
	\begin{split}
	\mu_F&=\alpha\mu \qquad \sigma^2_F=\alpha\sigma^2,\\
	\mu_M&=\beta\mu  \qquad \sigma^2_M=\beta\sigma^2,
	\end{split}	
	\end{equation}
	where $\mu$ and $\sigma^2$ are the mean and variance of $W$. 
	The variance-to-mean ratios of the variables $F$ and $M$ are equal to $\frac{\sigma^2}{\mu}$, which is the variance-to-mean ratio of $W$.
	Therefore, if $W$ is over-dispersed, $F$ and $M$ will also be over-dispersed.
	If we compare these variance-to-mean ratios with the case of the dependent variables (see \eqref{vmr}), we can see that the independent variables show a greater over-dispersion when $W$ is over-dispersed.
	

 	We now present an expression for each of the variables studied in the section \ref{sec:probapareamiento}. The proofs of these expressions can be found in the appendix \ref{formulasind}. 
 	Where $m$ is the mean of $W$ and $p_M$ is the probability mass function of $M$.
	\begin{itemize}
		\item Mean number of fertilized female parasites
		\begin{align}
		\alpha m \left[1-p_M(0) \right] 
		\end{align}
		
		\item Mating probability 
		\begin{align}
		\Phi=1-p_M(0) 
		\end{align}
		
		\item Mean egg production per host
		\begin{equation}
		\lambda_0G_M(z)G'_F(z)
		\end{equation}
		
		\item Mean fertilized egg production per host
		\begin{equation}
		\lambda_0 G_M(z) G'_F(z)\left[ 1-\frac{p_M(0)}{G_M(z)}\right]
		\end{equation}
	
		\item Mating probability with density-dependent effects
		\begin{equation}\label{phi-ind}
		\phi= 1-\frac{p_M(0)}{G_M(z)}
		\end{equation}
		
		\item Mean effective transmission contribution by female parasite
		\begin{equation}\label{psi-ind}
		\psi=\frac{G_M(z)G'_F(z)}{\alpha m}
		\end{equation}

		\item Contribution of mean fertilized egg production for mean-based deterministic model  of parasite burden
		 \begin{equation}
		\lambda_0 \alpha m \psi(m) \phi(m)
		\end{equation}
	\end{itemize}

\subsection{Some examples}
\label{examples-ind}
	In this section, we show the results obtained for some of the statistical models used in the section \ref{ss:examples-dep}.
%
	\subsubsection{Poisson}
	For the case where the distribution of parasites per host is Poisson with mean $m$, $W\sim \mathrm{Po}(m)$. A solution for the independence of variables $F$ and $M$ are the following distributions
	\begin{equation*}
	F\sim \mathrm{Po}(\alpha m), \qquad M\sim \mathrm{Po}(\beta m),
	\end{equation*}
	\begin{align*}
	G_F(s)G_M(s)&=e^{\alpha m(s-1)}e^{\beta m(s-1)}\\
	&=e^{m(s-1)}\\
	&=G_W(s).
	\end{align*}
	Note that the pgf of $F$ and $M$ coincide with what was obtained in section \ref{sec:distsexo}, which shows the independence of these variables in that section.
	The expressions obtained for $\psi$ and $\phi$ for this case are:
	\begin{equation}
	\begin{split}
		\psi(m)=e^{-m(1-z)},\qquad
		\phi(m)=1-e^{- m  z \beta}.
	\end{split}
	\end{equation}
	Note that the expression for $\psi$ and $\phi$ are the same as those obtained in the section \ref{ss:po-dep}.
	
	\subsubsection{Negative binomial}
	\label{nb-ind}
	For the case of a negative binomial parasite distribution with parameters $m$ and $k$,
	$W\sim \mathrm{NB}(m,k)$. 
	A solution for the independence of $F$ and $M$ are the following distributions:
	\begin{equation*}
	F\sim \mathrm{NB}(\alpha m,\alpha k), \qquad M\sim \mathrm{NB}(\beta m,\beta k),
	\end{equation*}  
	\begin{align*}
	G_F(s)G_M(s)&=\left[ 1-\frac{\alpha m}{\alpha k}(s-1)\right] ^{-\alpha k} \left[ 1-\frac{\beta m}{\beta k}(s-1)\right] ^{-\beta k}\\
	&=\left[ 1- \frac{m}{k}(s-1) \right] ^{-k}\\
	&=G_W(s).
	\end{align*}
	For this case, the pgf of $F$ and $M$ are not equal to those obtained in section \ref{sec:distsexo}, since it was shown that the variables were not independent. 
	The expressions obtained for $\psi$ and $\phi$ for this case are:	
	\begin{equation}\label{psinb-ind}
		\psi(m)=\left[ 1-\frac{m}{k}(z-1)\right] ^{-(k+1)},\qquad
		\phi(m)=1-\left[ \frac{1+\frac{m}{k}}{1-\frac{m}{k}(z-1)}\right]^{-\beta k} .
	\end{equation}
%
	Note that the expression $\psi$ is the same one obtained in the section \ref{ss:nb-dep}.
	On the other hand, the mating probability is $\Phi=1-\left( 1+\frac{m}{k}\right)^{-\beta k}$, which is the expression that \citet{may1977togetherness} obtained for $\beta=1/2$ and $W\sim \mathrm{NB}(m,2k)$.


	\subsubsection{Zero-inflated negative binomial}
	
	We now consider the case where the random variables $F$ and $M$ are zero-inflated negative binomial distributed.
	\begin{equation*}
		F\sim \mathrm{ZINB}\left( \pi,\tfrac{\alpha m}{1-\pi},\alpha k\right),  \qquad M\sim \mathrm{ZINB}\left( \pi,\tfrac{\beta m}{1-\pi},\beta k\right). 
	\end{equation*}  
	Note that $W$ is not distributed as a zero-inflated negative binomial. However, the mean $\mu$ and variance $\sigma^2$ of $W$ are
	\begin{equation}
		\mu=m,\qquad \sigma^2=m+\tfrac{m^2}{(1-\pi)k} + \pi m,
	\end{equation}	
%
	that coincide with the first moments of the model $\mathrm{ZINB}(\pi,\tfrac{m}{1-\pi},k)$ presented in section \ref{ss:zim-dep} for the case dependent variables.
	The expressions obtained for $\psi$ and $\phi$ for this case are: 
	\begin{equation}{\small
	\begin{split} 
		\psi(m)
			&=
			\pi \left[ 1-\tfrac{m}{(1-\pi) k}(z-1)\right] ^{-(\alpha k+1)}
			+ (1-\pi) \left[ 1-\tfrac{m}{(1-\pi) k}(z-1)\right] ^{-(k+1)},\\
			\phi(m)
			&=1- \frac{\pi + (1-\pi) \left[ 1+\frac{m}{(1-\pi) k}\right] ^{-\beta k}}{\pi + (1-\pi) \left[ 1-\frac{m}{(1-\pi) k}(z-1)\right] ^{-\beta k}}. 
	\end{split}
	}
	\end{equation}
%
	Note that the expressions for $\psi$ and $\phi$ are different from those obtained in the section on dependent variables \ref{ss:zim-dep}.
	The mating probability expression is given by
	$\Phi(m)=(1-\pi) \left\lbrace 1-\left [ 1+\frac{m}{(1-\pi) k}\right] ^{-\beta k} \right\rbrace$.
	The last expression is the mating probability $\Phi$ associated with $W\sim \mathrm{NB}(\tfrac{m}{1-\pi},k)$ as in the section \ref{nb-ind}, multiplied by the term $1-\pi$, which is the probability of observing non-zero values in the zero-inflated model.
	
\begin{figure}[h!]
		\centering 	\includegraphics[width=0.5\linewidth]{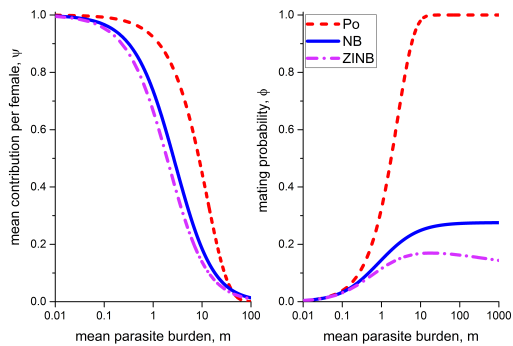}
		\caption{The mean effective contribution per female parasite, $\psi$ (left) and the mating probability, $\phi$ (right) corresponding to Poisson (dash curve), negative binomial (solid curve) and  zero-inflated negative binomial (dot curve) distributions. All as a function of the mean parasite burden $m$.}
		\label{f:psiandphi-ind}
	\end{figure}

	Plots of the effective transmission contribution ($\psi$) and the mating probability ($\phi$) for all the distributions discussed above are shown in Figure \ref{f:psiandphi-ind}. 
	We consider the parameters $\alpha=0.574$, $k=0.3$, $z=0.92$, $\pi=0.3$ \citep{anderson2014coverage,seo1979egg}.
	

\section{Monte Carlo simulations}

Monte Carlo simulations are computational algorithms that use random sampling and statistical analysis to solve problems or simulate complex systems.
A model is created to represent the real-world system, which is run many times with random inputs. The results are statistically analyzed to generate a distribution of possible outcomes.


%
%


\subsection{Model assumptions}
Simulation algorithms presented in this section are based on
the following assumptions and rules:

\begin{itemize}
	\item We considered a host population of size $N$.
	
	\item 
	The parasite burden of each host is a random variable $W$. Female and male burdens are random variables, $F$ and $M$ respectively. The values of the random variables will be denoted by $w$, $x$ and $y$ respectively. 

	\item Distributed together:
	The individual parasite burden per host is a value $w$ associated to the variable W with some assumed distribution (Poisson, negative binomial, etc.). 
	The individual female parasite burden ($x$) is then obtained as the result of $w$ Bernoulli trials with the success parameter $\alpha$, where $\alpha$ is the sex ratio of the female parasites. 
	The individual male parasite burden is then obtained as $y=w-x$.
	 
	
	\item Distributed separately:
	The values of the individual female and male parasite burdens ($x$ and $y$, respectively) are associated with the variables $F$ and $M$, respectively. These variables have assumed distributions (Poisson, negative binomial, etc.) that satisfy the equations \eqref{independencia}\eqref{indep-moments}.
	The individual parasite burden per host, $w$, is then obtained as $w= x + y$.
	



	\item The egg production per host is given by 
	$xz^{w-1}$,
	where $z=e^{-\gamma}$ with 	$\gamma$ the density-dependence intensity. 
	
	\item The infective egg production per host is given by 
	$I(y\neq 0)xz^{w-1}$,
	where $I$ is the indicator function of the set $y\neq 0$. 
	
	\item The fertilized female parasites are female parasites where the individual male parasite burden is non-zero ($y\neq 0$). 
	
	\item The mean effective transmission contribution per female parasite is obtained by quotient of the mean number egg production per host and the mean of female parasites per host.  
	
	\item The mating probability is obtained by quotient of the mean number of fertilized female parasites per host and the mean number of female parasites per host.
	
	\item The mating probabillity with density-dependent effects is obtained by quotient of the mean number infective egg production and the mean number egg production.


\end{itemize}

All the simulations were carried out in RStudio (Version 2022.12.0+353).

\subsection{Some examples}

In this section, we present the results obtained from Monte Carlo simulations, considering a distribution of parasites in the host population based on negative binomial and zero-inflated negative binomial models.

\subsubsection{Negative Binomial}


In this section, we report the empirical values of $\psi$ and $\phi$ obtained from Monte Carlo simulations of a host population of size N. 
For each host, we simulated its parasite burden based on a negative binomial model, and we 
distributed the parasites by sex according to their sexual radius. 

\begin{figure}[h!]
	\centering 	\includegraphics[width=0.5\linewidth]{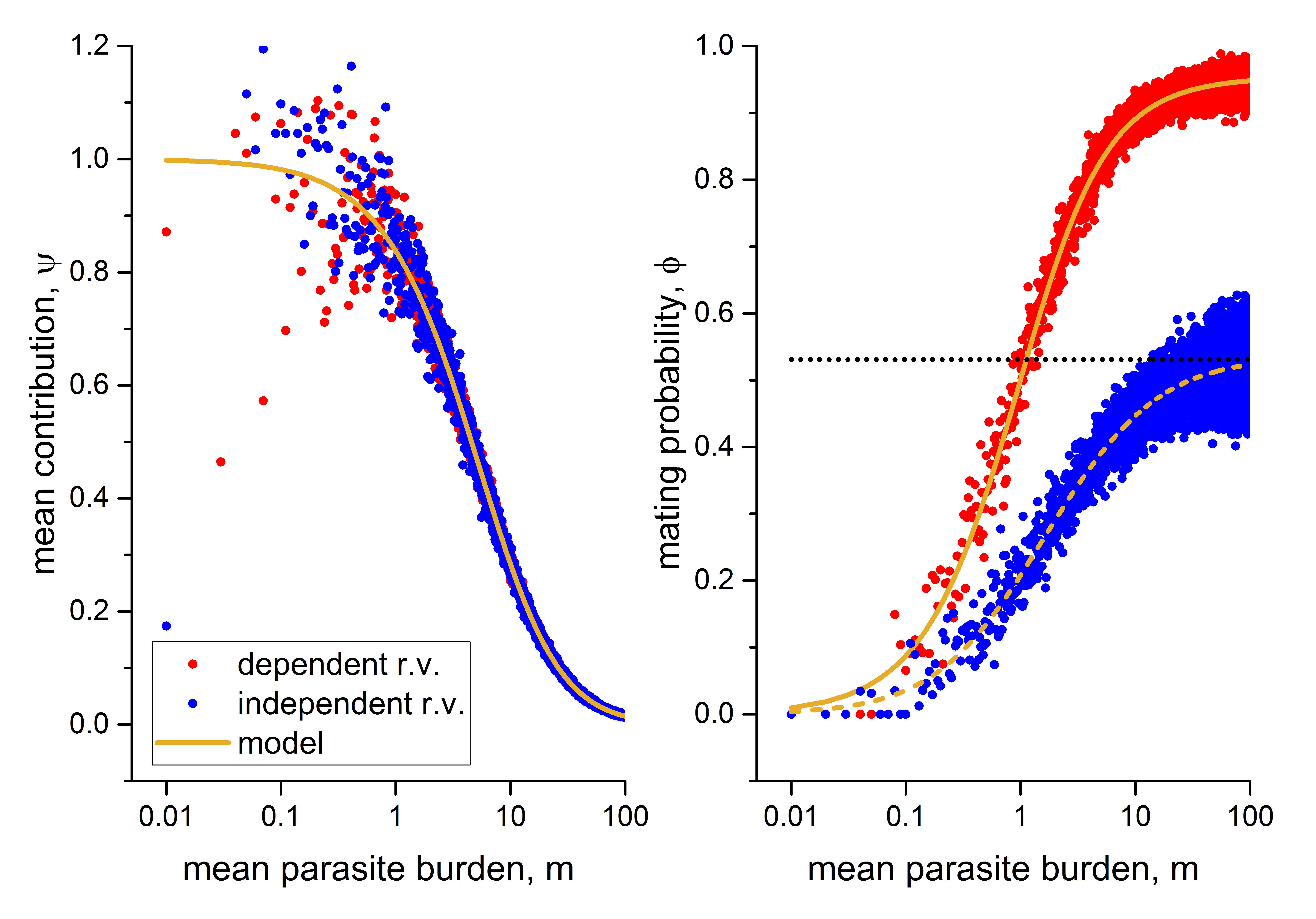}
	\caption{
		The empirical values (red and blue dots) of the mean contribution, $\psi$, and the mating probability, $\phi$, as a function of the mean parasite burden, $m$, for a parasite population with a negative binomial distribution. 
		The sample data were obtained by Monte Carlo simulations of a host population.
		We consider the parameters $\alpha=0.574$,  $k=0.7$, $z=0.92$  (\cite{anderson2014coverage,seo1979egg}).
	} 
	\label{f:simu-nb}
\end{figure}

Figure \ref{f:simu-nb} shows the empirical values of the mean contribution, $\psi$, and the mating probability, $\phi$, obtained from the Monte Carlo simulations. 
The blue and red points represent the empirical values of $\psi$ and $\phi$ for female and male parasites distributed together or separately, while the continuous and dashed curves show the theoretical models obtained for $\psi$ and $\phi$ respectively.	
	
%

As seen in Figure \ref{f:simu-nb}, the empirical values of $\psi$ are well-modeled by the theoretical model, as both models coincide for the cases of parasites distributed together or separately (as obtained in \eqref{phinb-dep} and \eqref{psinb-ind}). 


Similarly, the empirical values of $\phi$ for the cases of parasites distributed together or separately (blue and red points, respectively) are well-modeled by their respective theoretical models (continuous and dashed curves, respectively). 

We also show the asymptotes (black dotted lines) for each of the theoretical models obtained for $\phi$. It is worth noting that the values of the mating probability $\phi$ are close to one for the parasites distributed together case and further from one for the parasites distributed separately case.

Intuitively, we understand that the overdispersion in the distribution of parasites per host affects the mating probability since there are few hosts with many parasites, both females and males according to their sexual radius, to ensure mating. 
In the parasites distributed separately case, the negative effect of overdispersion is even greater because there are few hosts with many male parasites and few hosts with many female parasites. As a result, the occurrence of these two conditions in the same host is even more unlikely.

\section{Discussion and Conclusions}

We propose a model for the distribution of parasites among hosts, focusing on the distributions of females and males. Our model takes into account various reproductive variables of the parasites, such as the average number of fertilized female parasites, mean egg production, mating probability, mean fertilized egg production, and density-dependence effects.

We demonstrate that these reproductive variables are influenced by the independent nature of the female ($F$) and male ($M$) variables, as well as the density-dependent fecundity of the parasites. Interestingly, the reproductive expressions derived in our examples align with those found in previous works \citep{may1977togetherness,may1993biased,bradley1978consequences}. However, these earlier studies did not consider the impact of density-dependent fertility on the reproductive behavior of parasites.

The expressions we obtained serve as a generalization of the findings in \citep{may1977togetherness,may1993biased,bradley1978consequences}. It is important to note that our work focuses solely on parasites with a polygamous mating system, excluding monogamous and hermaphroditic parasites from consideration.

In conclusion, our study provides a comprehensive expression for the egg production and mating probability of parasites. We demonstrate how these expressions are influenced by the sex distribution of parasites and whether these distributions are considered joint or independent. Furthermore, we highlight the significant impact of density-dependence on parasite reproduction.

One of the main limitations of this work is that it only considers parasites with a polygamous mating system and we do not consider monogamous and hermaphroditic parasites.

In conclusion, in this work we obtained a general expression for egg production and the mating probability of the parasites. We show how these expressions depend on the sex distribution of the parasites and whether these distributions are considered joint or independent. 
We also show that these expressions vary due to the effects of the density-dependence of the parasites.

	\section*{Aknowledgements}
	
	This work was partially supported by grant CIUNSA 2018-2467. JPA is a member of the CONICET. GML is a doctoral fellow of CONICET.
	
	\section*{Conflict of Interest}
	
	The authors have declared no conflict of interest.

	\renewcommand{\bibname}{References}
	\bibliographystyle{apa}
	\bibliography{references}	

\end{document}